\documentclass[12pt]{article}

\usepackage{natbib,url}
\usepackage[german,greek,english]{babel}
\usepackage{alphabeta}
\usepackage{amsmath,amsthm}

\include{ProbBib,ThermBib,QBib}

\newcommand{\dbar}{d\mkern-6mu\mathchar'26}
\newcommand{\ex}[1]{\langle #1 \rangle}

\newcommand{\mc}{\mathcal}
\newcommand{\therm}{$\Theta \Delta^{cs}\;$}
\newcommand{\thrm}{$\Theta \Delta^{cs}$}
\newtheorem{proposition}{Proposition}
\newtheorem{corollary}{Corollary}[proposition]

\newcommand{\bibsuffix}[1]{}

\title{The Science of \thrm}
\author{Wayne C. Myrvold \\ Department of Philosophy \\ The University of Western Ontario \\ wmyrvold@uwo.ca}

\begin{document}
\maketitle
\abstract{There is a long tradition of thinking of thermodynamics, not as a theory of fundamental physics (or even a candidate theory of fundamental physics), but as a theory of how manipulations of a physical system may be used to obtain desired effects, such as mechanical work.  On this view, the basic concepts of thermodynamics, heat and work, and with them, the concept of entropy, are relative to  a class of envisaged manipulations.   This view has been dismissed by many philosophers of physics, in my opinion too hastily. This paper is a sketch and defense of a science  of  manipulations and their effects on physical systems. This is, I claim, the best way to make sense of thermodynamics as it is found in textbooks and as it is  practiced.  I  call this science \emph{thermo-dynamics} (with hyphen), or \thrm, for short, to highlight that it may be different from the science of thermodynamics, as the reader conceives it. Even if one is not convinced that it is the best way to make sense of thermodynamics as it is practiced, it should be non-controversial that \therm is a legitimate science. An upshot of the discussion is a clarification of the roles of the Gibbs and von Neumann entropies. Given the definition of statistical thermo-dynamic entropy, it can be proven that, under the assumption of availability of thermodynamically  reversible processes, these functions are the unique (up to an additive constant) functions that represent thermo-dynamic entropy. Light is also shed on the use of coarse-grained entropies.}
\newpage
\tableofcontents
\newpage
\section{Introduction} In what follows I will  tell you about a science that I call \emph{thermo-dynamics}.  Following the word of the Lord, when he first  bestowed that word upon us, I retain the hyphen, to emphasize the etymology of the word: it is formed from the Greek works for \emph{heat} and \emph{power}.\footnote{The word's first appearance is in Part VI  of Kelvin's ``On the Dynamical Theory of Heat'' (\citealt{KelvinDynamicalTheoryVI}, read before the Royal Society of Edinburgh on May 1, 1854). There he recapitulates what in \citeyear{KelvinDynamicalTheory} he had called the ``Fundamental Principles in the Theory of the Motive Power of Heat,''  now re-labelled  ``Fundamental Principles of General Thermo-dynamics.'' }  Following the word of the Laird, I will often abbreviate it as \therm (to be pronounced ``thermo-dynamics''), which also emphasizes its Greek roots.\footnote{Maxwell used this and related abbreviations in his correspondence with P. G. Tait. See  letter to Tait of Dec. 1, 1873, in  \citet[p. 947]{MaxLPII}.} The reason I  emphasize the etymology is that the science of thermo-dynamics has at its core a distinction between two modes of energy transfer between physical systems: as heat, and as work.

The concepts of \therm are, I claim, the best way to make sense of most of what is called ``thermodynamics'' in the textbooks, though that content is often obscured in the presentation.  Be warned, however: the scope of \therm is narrower than thermodynamics as it is sometimes conceived.  The scope of \therm   includes the zeroth, first, second, and (in the quantum context) third laws of thermodynamics, all of which were designated laws of thermodynamics by 1914 at the latest.  It does not include a relative late-comer to the family of laws of thermodynamics, which \citet{BrownUff} have dubbed the \emph{Minus First Law}, which, though it had long been identified as an important principle, was not called by anyone a law of thermodynamics prior to  the 1960s.\footnote{The minus first law, which Brown and Uffink also call  the \emph{Equilibrium Principle}, is given by them as,
\begin{quote}
An isolated system in an arbitrary initial state within a finite fixed volume will
spontaneously attain a unique state of equilibrium \citep[p. 528]{BrownUff}.
\end{quote}
They  point out that a principle of this sort had been recognized as a law of thermodynamics earlier, by Uhlenbeck and Ford (\citeyear[p. 5]{UhlenbeckFord}).}

A thermo-dynamic theory involves treating certain variables as \emph{manipulable}, in a sense that I will explain in the next section, and has to do with the responses of physical systems to manipulations of those variables.  A designation of certain variables as manipulable  is not something that appears in, or supervenes on, fundamental physics; it must be added. For this reason, \therm is not and cannot be a comprehensive or fundamental physical theory.  It is nonetheless a perfectly respectable theory, a useful one, and, for beings such as us, who are not transcendent intellects beholding the cosmos from outside but rather agents embedded in the world and interacting with it, perhaps even an indispensable theory.  Confusion arises when it is mistaken for the sort of theory that could possibly be a fundamental one.  Indeed, I will argue that some of the various puzzles and paradoxes that have arisen from thermodynamics stem from confusing \therm with fundamental physics.

The idea that thermodynamics should be thought of as a theory of this sort is not new; see Appendix for a sampling of quotations from the history of \thrm. A conception of thermodynamics along these lines is rapidly becoming the mainstream view among workers in quantum thermodynamics, who view it as a species of \emph{resource theory}, akin to quantum information theory (see \citealt{PRResourceTheories,TopicalReview,NgWoodsRTQT,Lostaglio2019}).  This development has so far attracted little attention from philosophers; a notable exception is \citet{WallaceControl}.

I start by outlining the basic concepts needed to formulate \thrm.

\section{Exogenous and manipulable variables}
In this section I highlight some routine features of scientific practice that are so ubiquitous that for the most part we don't really think about them, and they are passed over without comment.  It is worthwhile, however, to get clear about what's going on.

Consider a sort of problem that is frequently found in textbooks and in the scientific literature.  One is asked to consider a system subjected to an external force, or to an external potential, and to calculate certain aspects of its behaviour (\emph{e.g.} to  solve the equations of motion, or to find the energy eigenvalues), subject to that external potential. The Hamiltonian for such a system consists of its internal Hamiltonian, which includes the kinetic energies of its parts and terms involving interactions, if any, between its parts, plus the external potential.
\begin{equation}
H = H_{int} + V_{ext}.
\end{equation}
For the purposes of a problem like that, nothing needs to be said about the source of the external potential.  It is treated as \emph{given}.  Presumably,  the external potential is an interaction potential between the system in question and some other system, but we are not asked to include that system in our calculations, and, in particular, \emph{we do not consider the effect of the system under consideration on the system that is the source of the external potential}.  This is what it means to treat the external potential as given.

I will call variables treated in this way \emph{exogenous variables}.  Note that designation of a variable as exogenous has to do with how it is handled in a given investigation; the distinction between exogenous and other variables is not intrinsic to the physical nature of those variables.  The phrase ``exogenous variable'' should be taken as short for ``variable treated exogenously.'' Were it not for the awkwardness of language that would ensue, I would eschew adjectival uses of ``exogenous'' in favour of  adverbial.

The same variable may be treated exogenously in one investigation, and included as part of the system under consideration in another. An an example consider the usual pedagogical entry into celestial mechanics.  First one treats of a body in a fixed external $1/r$ potential, and shows that its trajectories take the form of conic sections (ellipses, parabolas, or hyperbolas, depending on the energy), subject to the area law with respect to the origin of coordinates.  This yields a respectable first approximation to planetary motion, as the gravitational effect of the sun dominates the net force on any planet, and, to a first approximation, the effect of the planet on the sun is negligible, and we may treat the sun as fixed.  The next step on the journey to celestial mechanics is from the one-body to the two-body problem, in which the sun's position is treated as dynamical variable.

We find this distinction between exogenous and other variables also in computer modelling of physical systems.  Consider climate models.  Various aspects of the earth's climate system are treated in such a model, and their behaviour subjected to the dynamics written into the model.  Some variables, such as solar radiation and greenhouse gas emissions from volcanoes  and from anthropogenic sources, are treated as inputs. No attempt is made to include solar dynamics or the geophysics of volcano eruptions in the dynamics of the model.

When dealing with exogenous variables, there is often a range of possible values to be considered, and we may be interested in the differences that changes to the exogenous variable make to the behaviour of the system at hand.  Crucially, we treat the exogenous variables as ones that can vary independently of the states of the systems under consideration---that is, they are treated as \emph{free variables}.  This is a crucial aspect of controlled experiments.  The systems under consideration are subjected to a range of treatments, and a well-designed experiment is one in which the treatments may be regarded as varying independently of the initial states of the systems to be studied.  Often some randomizing device is employed, which is thought of as rendering its outputs \emph{effectively free for the purpose at hand}.\footnote{Borrowing the apt phrase of \citet{FVLC}.}

I will say that a  variable is being treated as a \emph{manipulable} variable in a given theoretical investigation if (i) it is being treated as exogenous variable, and (ii) there is a range of possible values, or, perhaps, of possible alternative temporal evolutions of that variable, under consideration.

One might be tempted to say that treatment of certain variables as exogenous is a concession to our limited calculational and computational abilities.  It might be better, one might think, to include in our climate models solar dynamics, the dynamics of volcanoes, and a sufficiently detailed model of human activities that anthropogenic emissions could be included among the modelled variables.  This would be a mistake.  For certain purposes, it is \emph{essential} to treat certain variables as manipulable.  These purposes include attribution studies.  To use climate models to estimate the contribution of various inputs to observed global warming, researchers vary those inputs while holding others fixed.  It is investigations such as these that, in part, underwrite conclusions that most, all, or perhaps more than all of the observed warming can be attributed to anthropogenic greenhouse gases.  And, of course, this is relevant to policy decisions (or would be, if anyone were making informed policy decisions); one can make projections by modelling future climate under a variety of emissions scenarios.

All of this is, of course, meant to be consistent with the concept of manipulability as it appears in the causal modelling literature \citep{PearlCausality,WoodwardMaking,WoodwardSEP}.

When speaking of manipulable variables, and a set of alternative manipulations, one almost inevitably begins to talk of \emph{choices} of manipulations. This carries with it a suggestion that human agency is central to the concept, which in turn raises the suspicion that subjectivity is being brought in. This is not the case; a variable treated as manipulable need not be manipulable by \emph{us} (see above, re volcanoes).  Nevertheless,  some  who have developed a conception very close to what I am calling \therm  have lapsed into talk that suggests that its concepts are subjective. This is an error, in my view.  It stems, I think, from overextension of the familiar subjective/objective dichotomy.  Objective features of a physical system are supposed to belong to that system, in and of itself; they are features that cannot change without change of its physical state.  The concepts of \therm are relative to a specification of manipulable variables and a set of alternative manipulations of those variables, and as such are not there in the physical states of things.  It does not follow that they are subjective, although, if all one had at hand was the objective/subjective dichotomy, it is understandable that one might lapse into saying that they are.

\section{Thermo-dynamic theories}\label{EQTD}  An equilibrium  thermo-dynamic state of a system $A$ may be specified by its total internal energy $E$ and the values of one or more manipulable variables $\boldsymbol{\lambda} = \{ \lambda_1, \ldots, \lambda_n\}$.  As a  running example, you can think of a gas confined to a container with a moveable piston, whose walls are represented as an external potential that strongly repels molecules that get too close. We consider a family of such potentials, corresponding to different positions of the piston.

It is often assumed that, besides changes to the variables $\boldsymbol{\lambda}$, there are other manipulations that may be performed.  For example, the system $A$ may be coupled to other systems regarded as heat reservoirs at various temperatures.  This coupling may be applied or removed; that is, the interaction Hamiltonian between $A$ and the heat reservoir is being treated as a manipulable variable. A heat reservoir is a system with which is associated a definite temperature, from which no work is extracted and on which no work is done; its only exchanges of energy with other systems are as heat.  What it means to count a system as a heat reservoir at a given temperature will be discussed a bit more in the next section.  Often, one imagines heat reservoirs available for arbitrary temperatures.  But one can also consider the thermo-dynamic theory of an adiabatically isolated system, or a theory on which there is access to only one heat reservoir, or some other limited set.

Corresponding to any manipulation is a transformation of the state of the system.   A small change $d\lambda_i$ in one of the manipulable variables, with the others held fixed, and no heat exchange, may result in a change $dE$ in the internal energy of the system.  We define,
 \begin{equation}\label{Ai}
 A_i(E, \boldsymbol{\lambda}) =  \frac{\partial E}{\partial \lambda_i},
\end{equation}
where it is understood that the other variables are held fixed, and there is no exchange of energy with  any heat reservoir or anything else. In standard thermodynamics, the quantities $A_i$  are usually assumed to have steady, time-independent values.  We can take this condition (which will be modified in section \ref{StatTD}) as a criterion of thermal equilibrium of the system.  In any process involving a small change in the variables $\boldsymbol{\lambda}$, we define work done on the system as
\begin{equation}
\dbar W = \sum_i A_i \, d \lambda_i.
\end{equation}
The convention in play is that work done \emph{on} the system, increasing its energy, counts as positive.   If the only other changes to the internal energy of the system $A$ are due to interactions with heat reservoirs, we have a neat partitioning of any change in the energy of $A$ into a work component and a heat component.  Changes in energy of $A$ due to changes in the manipulable variables counts as work; exchanges of energy with heat reservoirs, as heat.  As with work, we count heat transfer into the system $A$ as positive.

A thermo-dynamic theory consists of a system $A$, a class of Hamiltonians $H_{\boldsymbol{\lambda}}$ that depend on manipulable variables $\boldsymbol{\lambda}$, and a set $\mc{M}$ of possible manipulations of those variables.  The class might include manipulations that go beyond what can feasibly be achieved by us; we can very well consider how a system would react to more fine-grained manipulations than we can achieve, or to manipulations that proceed so slowly that we would not have the patience to see them through. What one needs to know about the effects of these manipulations is given by the dependence of the generalized forces $A_i$ on the values of the parameters $(E, {\boldsymbol{\lambda}})$ specifying the state.  The structure of the set of manipulations may vary from theory to theory.  One thing that I will assume in what follows is that manipulations can be composed: if there is a manipulation that takes a state $a$ to a state $b$, and a manipulation that takes a state $b$ to a state $c$, these manipulations can be performed in succession, forming a manipulation that takes state $a$ to $b$ and then to $c$.

We will not be assuming that thermodynamically reversible processes, or even processes that approximate thermodynamic reversibility arbitrarily closely, are always available.  Dropping the assumption of the availability of reversible processes requires revision of the familiar framework of thermodynamics, as it means dropping the assumption of the availability of an entropy function.  In its place we will define quantities $S_\mc{M}(a \rightarrow b)$, defined relative to a class of available manipulations $\mc{M}$, to be thought of as analogues, in the current context, of entropy difference between states $a$ and $b$. These will be representable as differences in the values of some state function \emph{only} in the limiting case in which all states can be connected reversibly.

For any two thermo-dynamic states $a$, $b$, let $\mc{M}(a \rightarrow b)$ be the set of manipulations in $\mc{M}$ that lead from $a$ to $b$. These may involve heat exchanges with one or more heat reservoirs $\{B_i \}$ with temperatures $T_i$.  For any manipulation $M$ in $\mc{M}(a \rightarrow b)$, let  $Q_i(a \rightarrow b)_M$ be the heat transferred over the course of $M$ into $A$ from the reservoir $B_i$ (positive if there is energy flow from $B_i$ to $A$, negative if there is energy flow the other way).  We define,
\begin{equation}
\sigma_M(a \rightarrow b)  = \sum_i \frac{Q_i(a \rightarrow b)_M}{T_i}.
\end{equation}
 We define, as analogues of entropies (which we will henceforth just call ``entropies''),\footnote{The word, appropriately, is formed from the Greek \textgreek{\<\eta  \enspace \tau\rho\omega\pi\`{\eta}},
transformation, for what Clausius called the \emph{transformational content} (\emph{Verwandlungsinhalt}) of a body \citep[p. 390]{Clausius65}.}
\begin{equation}\label{entropy}
S_\mc{M}(a \rightarrow b) = \mbox{l.u.b.} \: \{ \sigma_M(a \rightarrow b)  \, | \, M \in \mc{M}(a \rightarrow b) \}.
\end{equation}
where ``l.u.b.'' stands for ``least upper bound,'' that is, the smallest real number that is at least as large as all members of the set.\footnote{If $b$ cannot be reached from $a$ via any manipulation in $\mc{M}$, or if the set considered has no upper bound, $S_\mc{M}(a \rightarrow b)$ is undefined.  To avoid qualifying every formula involving entropies with a {proviso}  that all quantities mentioned therein are defined, we can, if we like, allow $S_\mc{M}(a \rightarrow b)$ to take values in the extended reals, which supplement the reals with   $\pm \infty$. Then, if $b$ cannot be reached from $a$, $S_\mc{M}(a \rightarrow b) = -\infty$.} Via the obvious extension of this definition we also define quantities such as $S_\mc{M}(a \rightarrow b \rightarrow c)$ for processes with any number of intermediate steps.   It follows from the assumption about  composition of manipulations and the definition of the entropies that
\begin{equation}
S_\mc{M}(a \rightarrow b \rightarrow c) = S_\mc{M}(a \rightarrow b) + S_\mc{M}(b \rightarrow c),
\end{equation}
and similarly for processes with longer chains of intermediate states.

One version of the second law of thermodynamics says that, if a system undergoes a cyclic transformation, returning it to its original state, the sum of $Q/T$ over all heat exchanges in the process cannot be positive.  We can write this as:
\begin{quote}
\emph{\textbf{The second law of thermo-dynamics}}. For any state $a$, \[S_\mc{M}(a \rightarrow a) \leq 0.\]
\end{quote}
It follows from the second law that, for any states $a$, $b$,
\begin{equation}
S_\mc{M}(a \rightarrow b \rightarrow a) \leq 0,
\end{equation}
and similarly for cycles  consisting of larger numbers of states.

By the second law, $S_\mc{M}(a \rightarrow b  \rightarrow a)$ cannot exceed zero.  If it is equal to zero, then there is no harm in adding to the list of possible manipulations a fictitious reversible process that can be run in either direction, from $a$ to $b$, or, with signs of heat exchanges reversed, from $b$ to $a$.  We don't expect any actual process to satisfy this condition; as John \citet{ImpossibleProcess} has emphasized, any process will involve some dissipation of energy, and fail to be completely reversible. If one took talk of reversible processes too literally, one would end up ascribing absurd properties to them; they would be processes that take place infinitely slowly and yet somehow manage to get completed.  Norton argues that talk of reversible processes should be regarded as short-hand for talk of limiting properties of sets of actual processes.  Our definition of  entropy makes this explicit.

In what follows, take the statement that $a$ and $b$ can be reversibly connected as no more than a convenient way of saying that $S_\mc{M}(a \rightarrow b \rightarrow a)$ is equal to $0$. On the macroscopic scale, it may be the case that, for all $a$, $b$, $S_\mc{M}(a \rightarrow b \rightarrow a)$ is close enough to zero that we can neglect the fact that it is not exactly zero. In standard thermodynamics, which is usually meant to apply at the macroscopic scale, it is normally assumed that {any} two states of a system can be connected by a reversible process.  If this holds---that is, if, for all states $a$, $b$, $S_\mc{M}(a \rightarrow b \rightarrow a)= 0$---it follows from the second law  that there is a function $S$ on the set of thermodynamic states,  defined up to an additive constant, such that
\begin{equation}
S_\mc{M}(a \rightarrow b) = S_\mc{M}(b) - S_\mc{M}(a).
\end{equation}
(Proof left as an exercise for the reader.)  If, however, we want to push \therm down to the nanoscale, on which departures from reversibility are non-negligible, we need not assume this.

Call a transformation from a thermo-dynamic state $a$ to a state $b$ \emph{adiabatic} if no exchanges of energy occur that are not due to manipulation of the variables ${\boldsymbol{\lambda}}$; no heat is exchanged with any heat reservoir.  The following is a simple consequence of the definition of  entropy.
\begin{proposition}\label{NoDemon}
If there is a manipulation that takes state $a$ to state $b$ adiabatically, then, for any state $c$, $S_\mc{M}(b \rightarrow c) \leq S_\mc{M}(a \rightarrow c)$  and $S_\mc{M}(c \rightarrow b) \geq S_\mc{M}(c \rightarrow a)$.
\end{proposition}
In the special case in which  all states are reversibly connectable, this says that an adiabatic transformation cannot lower the entropy of a state.

It's a consequence of all this that, given  a physical system $A$, there may be several thermo-dynamical theories of that system $A$, depending on the specification of manipulable variables, and on the set $\mc{M}$ of possible manipulations.  This means that a pair of physical states $a$, $b$ of the system might be assigned different values of the entropy $S_\mc{M}(a \rightarrow b)$ by different thermo-dynamic theories.  This will be illustrated in the next section.  If one thought that the entropy difference of a pair of states of  a system was supposed to be a property of those physical states alone,  this might seem paradoxical.  In the context of \thrm, there's nothing paradoxical about it at all.

Once the set $\mc{M}$ of possible manipulations  is chosen, how the system reacts to those manipulations is a matter of physics.  These reactions are encoded in the equilibrium values of the generalized forces $A_i$, defined by (\ref{Ai}).  It is these that determine the dependence of  entropies $S_\mc{M}(a \rightarrow b)$ on the values of the manipulable variables.  One may say: we may choose the variables to manipulate, but nature chooses the response to those manipulations. It would be mistake to say that a view of this sort makes entropy subjective.   Entropy remains a measurable quantity, but what quantity it is that is being measured is determined by the choice of manipulable variables.

What we have presented in this section is almost the same as what is found in typical thermodynamic textbooks.  Almost.  It is universally agreed that thermodynamic states are defined relative to some selection of  a set of variables that is small, compared to the full set of variables needed to specify the precise physical state of a system. The difference is that these variables are often described as the \emph{macroscopic} variables, the ones whose values can be obtained via a macroscopic measurement.

What to say about this?  First: though this is not always explicitly said, if one reads any textbook of thermodynamics closely enough, one will find that the extensive variables that define an equilibrium state are invariably treated as manipulable variables, in the sense discussed in the previous section.\footnote{Example, from one of the most widely used textbooks,
\begin{quote}
A description of a thermodynamic systems requires the specification of the ``walls'' that separate it from its surroundings and that provide its boundary conditions. It is by means of manipulations of the walls that the extensive parameters of the system are altered and  processes are initiated  \citep[p. 15]{CallenBook2nd}.
\end{quote}
}  Sometimes they are called \emph{external variables}.  Second: it should be stressed that the selected  variables are not properties of the system to be studied, but of external constraints placed on the system. For example, the quantity $V$ that appears in the equation of state of a gas is the volume \emph{available to} the gas. Third: even if there is a correspondence between manipulable variables and   macroscopic extensive variables (as there is a correspondence between the position of the walls of a container and the volume occupied by a gas in its equilibrium state),  these are conceptually distinct.  Fourth: an equilibrium thermo-dynamic state need not be a state in which all macroscopically observable quantities have stable values.  Consider, for example, a particle, visible under a microscope of modest power, undergoing Brownian motion.  If---as I think we should---we count its position as macroscopically observable, this does not settle down to a stable value.  What we have, instead, is a stable pattern of fluctuations.  This can well count as a state of thermo-dynamic equilibrium.

\section{Examples}  Two examples  will help illustrate how \therm works, and how it differs from the standard way of presenting thermodynamics.
\subsection{Entropy of mixing of gases}  Consider the following example, discussed by Gibbs (\citeyear[227--229]{GibbsEHSI}; \citeyear[166-167]{GibbsSciPapersI}),  which has been the topic of considerable discussion since that time.  We consider a container divided by a partition into two subvolumes, each containing samples of gas at the same temperature and pressure.  The partition is removed, and the gases interdiffuse, until each is equally distributed within the whole volume.  Has there been an increase of entropy, or not?

The  answer found in all the textbooks,  given already by Gibbs, is that if the gases initially in the two subvolumes are of the same type,  there has been no change of thermodynamic state, and \emph{ipso facto} no change in entropy.  If the two subvolumes initially contain gases of different types, initial and final states of the contents of the container are distinct thermodynamic states, and the entropy of the final state is higher than that of the initial state.       This entropy increase is known as the \emph{entropy of mixing}.

But what is the criterion for sameness of thermodynamic state?  On the standard textbook account, thermodynamic states are defined with respect to macroscopic variables.  On this account, initial and final states are distinct if and only if they macroscopically distinguishable.  On the thermo-dynamic account, initial and final states count as distinct only if the class $\mc{M}$  contains manipulations that act differentially on the two gases, in such a way that their interdiffusion represents a lost opportunity to extract work.  A standard textbook device, originating with Boltzmann (\citeyear{BoltzDiff}) and popularized by \citet[\S 236]{PlanckThermodynamik1}, involves conceiving of pistons made from some material permeable to one gas but not the other.  Armed with such pistons, one could  slowly expand one gas and then the other, keeping their temperature constant as work is extracted by having them in contact with a heat reservoir.  In such a way one obtains the standard entropy of mixing, which is just the sum of the entropies of  expansion of the two gases.

One could imagine cases in which the initial and final states are macroscopically distinct but not thermo-dynamically distinct.  They could, for example, differ in colour.  If the class of manipulations considered does not include any way to exploit this difference to differentially manipulate them, then initial and final states will not, for that thermo-dynamic theory, be distinct thermo-dynamic states.  Initial and final states could also be thermo-dynamically distinct but not macroscopically distinguishable via the sorts of operations we usually count as macroscopic observations.  They might appear the same to our measuring apparatus, and still react differently to the aforementioned semi-permeable pistons.

As a historical note: conflation of these two notions of thermodynamic state goes back as far as Gibbs' discussion, as Gibbs gives both answers to the question of criterion of distinctness of initial and final states.  He first gives, as a criterion for restoring the initial state of the gases,  the condition that we bring about a state ``undistinguishable from the previous one in its sensible properties'' (\citeyear[p. 228]{GibbsEHSI}; \citeyear[p. 166]{GibbsSciPapersI}).   ``It is to states of systems thus incompletely defined,'' he  says, ``that the problems of thermodynamics relate.''  But then, in the following paragraph, he writes,
\begin{quote}
We might also imagine the case of two gases which should be absolutely identical in all the properties (sensible and molecular) which come into play while they exist as gases either pure or mixed with each other, but which should differ in respect to the attractions between their atoms and the atoms of some other substances, and therefore in their tendency to combine with other substances. In the mixture of such gases by diffusion an increase of entropy would take place, although the process of mixture, dynamically considered, might be absolutely identical in its minutest details (even with respect to the precise path of each atom) with processes which might take place without any increase of entropy. In such respects, entropy stands strongly contrasted with energy.  (\citealt[pp. 228--229]{GibbsEHSI}; \citeyear[p. 167]{GibbsSciPapersI})
\end{quote}
Here he seems to be acknowledging that the key issue is not whether the two gases are the same in their sensible properties, but whether or not they can be separated by external means.

This example has  given rise to metaphysical discussions that are completely irrelevant.   The relevant criterion of distinctness, it is said in some quarters, is whether the particles of the two gases  are identical in a strong sense, according to which exchange of  particles  makes \emph{no difference whatsoever} to the physical state.  On such a view, if all the particles were distinct---that is, if every particle involved differed in some physical property from all the others---then there would always be an entropy of mixing when the barrier was removed.  As Robert Swendsen (\citeyear{SwendsenColloids,SwendsenGibbs}) has argued, this gives the wrong answer when applied to a colloidal suspension.  A colloid, such as paint, or milk, consists of blobs, called colloidal particles, of some type of material suspended in some fluid.  The colloidal particles may be large enough that each contains a large number of molecules, and, though their sizes may be sufficiently uniform that we are justified in treating the colloid as a collection of identical particles, it might be that no two of them contain exactly the same number of atoms. Someone committed to the position that for a collection of distinct  particles there is always an entropy of mixing when a partition is removed would be committed to the position that we can lower and raise the entropy of a can of paint merely by inserting or removing a partition.  This is the wrong answer.  In the absence of any means of manipulation that is so sensitive to the minute differences between colloidal particles that  each particle can be differentially manipulated, there is no entropy of mixing when one removes a partition separating two samples of the same type of paint.

The entropy of mixing of two distinct gases depends only on the quantities of gas in each subvolume, and on their initial and final volumes.  It is independent of the \emph{degree} of dissimilarity.  This struck Duhem (\citeyear{Duhem1892}) as paradoxical, and, following him, Wiedeberg (\citeyear{Wiedeberg1894}), who spoke  of ``Gibbs' paradox.''  The alleged paradox stems from a tension  between the independence of the entropy of mixing from the nature of the gases (as long as they are distinct), and the idea that a result valid for identical gases should be obtainable as a limit-case of distinct gases of diminishing degree of dissimilarity.

If entropy is thought of as an intrinsic property of a system, like its mass or its total energy, then this does seem puzzling.  However, as argued by  Denbigh and Redhead (\citeyear{DenbighRedhead}), if we recall how entropy is defined---relative to some set of processes, as a limit of some quantity taken over all processes in that set---this does not seem surprising at all.  The result of any particular  process, taking placing within a fixed duration of time, may well depend  continuously on the relevant parameters of the system.  But entropy involves a limit over a set of processes.  As two gases become more and more similar, the time required to achieve a given degree of separation may increase, but, if our set of manipulations contains arbitrarily slow processes, this will not  affect entropy as a limit property.

An analogy may help.  Consider a collection of immortal ants that crawl at different rates towards a hill that is one metre tall.  All of them, as long as they have a nonzero velocity in the proper direction, eventually reach the top of the hill.  The distance crawled, and height reached, at any given time $t$, is a continuous function of the speed at which the ant crawls.  But the maximum height reached by an ant is one metre for any nonzero speed, and zero for a stationary ant, and so is a discontinuous function of the ant's speed.

\subsection{Helmholtz free energy} Suppose that the class of manipulations to be considered involves access to only one heat reservoir, at temperature $T$.  We ask: if the system starts out in a state $a$ and ends up in state $b$, what is the most work that you can extract from it along the way?

Let $E_a$ and $E_b$ be the internal energy of the system in states $a$ and $b$, respectively.  If work is extracted from the system, this means that $W$ is negative.  We obtain from the system a positive amount of work $W_{gain} = -W$.  Conservation of energy requires,
\begin{equation}
E_b - E_a = Q - W_{gain}.
\end{equation}
From the definition of  entropy $S_\mc{M}(a \rightarrow b)$,
\begin{equation}
S_\mc{M}(a \rightarrow b) \geq \frac{Q}{T},
\end{equation}
and so
\begin{equation}\label{Wout}
W_{gain} \leq -(E_b - E_a - T S_\mc{M}(a \rightarrow b)).
\end{equation}
If the quantity on the right-hand side of (\ref{Wout}) is negative, then no work can be obtained in a transition from $a$ to $b$ using a heat reservoir at temperature $T$ as a resource; on the contrary, the transition requires expenditure of a quantity of work (that is, a positive quantity of energy going into the system),
\begin{equation}
W_{cost} \geq E_b - E_a - T S_\mc{M}(a \rightarrow b).
\end{equation}
Call the quantity
\begin{equation}
F_\mc{M}(a \rightarrow b) = E_b - E_a - T S_\mc{M}(a \rightarrow b)
\end{equation}
the \emph{Helmholtz free energy} of $b$ relative to $a$. If the only available heat sources and sinks are at temperature $T$,   a transition from $a$ to $b$ is achievable without expenditure of a positive  quantity of work if and only if $F_\mc{M}(a \rightarrow b) < 0$.

Let us now make the assumption that all states are reversibly connectible, and hence that there is a state-function $S$ available, such that $S_{\mc{M}}(a \rightarrow b) = S_\mc{M}(b) - S_\mc{M}(a)$.  This allows us to define a function
\begin{equation}\label{F}
F_\mc{M} = E  - T S_\mc{M}
\end{equation}
such that
\begin{equation}
F_\mc{M}(a \rightarrow b) = F_\mc{M}(b) - F_\mc{M}(a).
\end{equation}
The quantity $F_\mc{M}$ was called the \emph{available energy} in the 4th edition (\citeyear{MaxTOH4}), and subsequent editions, of Maxwell's \emph{Theory of Heat} (pp. 187--192).  It was called \emph{freie Energie} by \citet{Helmholtz1882}, whence its current name, \emph{Helmholtz free energy}. If all heat exchanges are with reservoirs at temperature $T$, then a transition from $a$ to $b$ requires work to be done if $F_\mc{M}(b) > F_\mc{M}(a)$, and can be a source of work if $F_\mc{M}(b) < F_\mc{M}(a)$.

There is an interesting difference between the uses of this concept by Maxwell and Helmholtz, respectively.  Helmholtz imagines a system in contact with a heat bath at temperature $T$.  All changes under such conditions are isothermal changes, and  the free energy difference between two states is the work needed to effect a state transition via an isothermal process.  The use of the concept is to determine the equilibrium state of the system, which is the state in which $F$ takes its minimum value (that is, work has to be done to move the system away from this state).  This is the use to which it is put in most modern textbooks.  This presentation may suggest that the Helmholtz free energy is  a property of the system itself.

Maxwell, on the other hand, imagines transitions between arbitrary initial and final states; these need not be states of temperature $T$.  The change in available energy is the work needed to effect a state transition, using a heat reservoir at temperature $T$ as a resource.  On this way of thinking about it, $F$ is a function both of the state of the system, via state functions $E$ and $S$, and of the heat reservoir, via $T$.

\section{Statistical thermo-dynamics}\label{StatTD} In the previous section it was assumed that the equilibrium values of the quantities $\{A_i\}$, defined by eq. (\ref{Ai}), are well-defined as functions of the energy $E$ and the manipulable variables ${\boldsymbol{\lambda}}$.

That this is a substantive assumption can be seen by considering the example of a gas confined to a container with a moveable piston whose position is taken to be  manipulable.  The generalized force corresponding to displacements of the piston is the negative of the pressure.  For  a macroscopic gas in equilibrium, we expect an even and steady pressure on the walls of the container.  If we think about what is happening on the molecular level, we realize that this is a statistical regularity of the same sort as the observed near-constancy of deaths per capita in a given population from year to year, a regularity arising from aggregation of  a large number of individually unpredictable events. A regularity of this sort is not to be thought of as something that occurs with certainty, but, rather, with high probability.  If we ask whether we could push on the piston and find ourselves able to diminish the volume with virtually no resistance, we have to admit that it is not impossible, but (for a macroscopic gas) so highly improbable that the possibility may be neglected.

This means that probabilistic considerations are in play, even in the cases where there is a determinate (enough) near-certain amount of work required for a given manipulation.  The role of probability may be left implicit in cases where deviation from certainty is negligible.  However, since probability is playing a role whether explicitly acknowledged or not, it is best to introduce probabilistic considerations explicitly.  This opens up the possibility of a more general theory that embraces cases in which statistical fluctuations in generalized forces are non-negligible, with  the quasi-deterministic macroscopic theory as a limiting case.

It is a commonplace of the literature on philosophy of probability that the word ``probability'' is used in more than one sense.  That raises the question of what probability is to mean in this context.  I will defer that question (but see \citealt{MyrvoldOxfordHandbookProb} for some options), leaving a gap in the account to be filled in.  As long as the usual machinery of probability theory is applicable, the conclusions we will draw will be independent of how that gap is filled.

One thing should be stressed, however. In the latter half of the nineteenth century, it became increasingly common (spurred, in part by Venn's \emph{The Logic of Chance}) \nocite{VennChance} to think of probability statements as involving veiled reference to frequencies in some actual or hypothetical series of similar events.  It was in this milieu that \citet{Boltzmann1871a,BoltzGasTheorieII}, and, following him, \citet{Maxwell1879b} and \citet{GibbsBook}, began to think in terms of an imaginary ensemble consisting of a large number of systems with the same external parameters and varying microstates.  Frequentism is widely (and rightly, in my opinion) rejected in the literature on the philosophy of probability. Fortunately, nothing in the approach of Boltzmann and his successors is wedded to it.  Any readers who have qualms about talk of probabilities stemming from a worry that probabilities cannot be ascribed to individual systems should rest assured that  this is not the case. There is no commitment to frequentism about probabilities.  Feel free to take the talk of ensembles by Boltzmann, Maxwell, Gibbs, and the textbook tradition that followed as a picturesque way of talking about a probability distribution applied to propositions about an individual system.

Given a thermo-dynamic state of a system, we want to have probability distributions over the work done and heat exchanged as a  result of a manipulation.  The reason that these don't have determinate values is that the thermo-dynamic state of a system drastically underspecifies the physical state of the system.  This suggests that we supplement our specification of a thermo-dynamic state, which so far involves specification of the internal energy and of values of the manipulable variables, with a specification of a probability distribution over possible physical states of the system.  This can be done in the context of  classical or quantum mechanics.  In a classical context,  we will have  assignments of probabilities to appropriate subsets of the system's phase space; in the quantum context,  probability distributions over the pure states of the system.

What now happens to the second law of thermodynamics?  In a regime in which statistical fluctuations of the force on a piston are non-negligible, we might in  a given cycle of an engine end up expending less work than expected in the compression stage, and hence might obtain in that cycle more work than the Carnot limit.  But, by the same token, we might expend more work than expected.  We  expect that we won't be able to \emph{consistently and reliably} violate the Carnot limit on efficiency.  This suggests a probabilistic version on the second law, expressed in terms of expectation values of heat and work transfers.  The second law will then be, to employ Szilard's vivid analogy, like a theorem about the impossibility of a gambling system intended to beat the odds set by a casino.
\begin{quote}
Consider somebody playing a thermodynamical gamble with the help of cyclic processes and with the intention of decreasing the entropy of the heat reservoirs. Nature will deal with him like a well established casino, in which it is possible to make an occasional win but for which no system exists ensuring the gambler a profit (\citealt[p. 73]{Szilard1925e}, from \citealt[p. 757]{Szilard1925}).
\end{quote}

We will be considering exchanges of energy with heat reservoirs.  A heat reservoir is a system from which no work is extracted and on which no work is done; its only exchanges of energy with other systems are as heat. When two heat reservoirs of the same temperature are placed in thermal contact, there is no tendency for heat to be transferred in either direction, and the expectation value of the heat exchange is zero. When two reservoirs are placed in thermal contact, the expectation value of heat flow is from warmer to cooler.  Any collection of heat reservoirs at the same temperature may be regarded as a larger heat reservoir at the same temperature.

From considerations of this sort one can argue (see \citealt{MaroneyEntropy} for exposition) that an appropriate probability distribution to associate with a heat reservoir is the one that Gibbs labelled the \emph{canonical distribution}.  In the classical context, it is defined as the distribution with density function, with respect to Liouville measure,
\begin{equation}
\tau_\beta(x) = Z^{-1} e^{- \beta H(x)},
\end{equation}
where $\beta$ is the inverse temperature $1/kT$, and $Z$ is the normalization constant required to make the integral of this density over all phase space unity.  This depends both on the Hamiltonian $H$ and on $\beta$, and is called the \emph{partition function}.  In the quantum context,  the canonical distribution is represented by a density operator,
\begin{equation}
\hat{\tau}_\beta = Z^{-1} e^{- \beta \hat{H}},
\end{equation}
where, again, $Z$ is the constant required to  normalize the state.  We will henceforth take it that to treat a system as a heat reservoir is to represent its thermo-dynamic state by a canonical distribution, uncorrelated with the rest of the world.

\section{Statistical-mechanical entropies, and the second law}  In the spirit of Szilard's analogy, if we seek a statistical-mechanical analog of the thermo-dynamic entropy, we may take the definition (\ref{entropy}) and replace the heat exchanges mentioned therein with their expectation values.

A thermo-dynamical state of a system will be specified by its Hamiltonian $H$, which  depends on manipulable variables $\boldsymbol{\lambda},$  together with  a probability distribution over its state space.  In the classical context the probability distribution may be represented by a density function $\rho$; in the quantum context, the salient aspects of such a distribution may be represented by a density operation $\hat{\rho}$.  Given a thermo-dynamical state $a = ( \rho_a, H_a )$, we consider the effects of some manipulation, which may consist of manipulation of the variables ${\boldsymbol{\lambda}}$ and of couplings to various heat reservoirs $\{B_i\}$.   The probability distribution for $A$, together with canonical distributions for the heat reservoirs, determines an initial probability distribution over the composite system consisting of $A$ and the reservoirs $\{B_i\}$.  This will evolve, in accordance with the Liouville equation (classical) or Schr\"odinger equation (quantum), according to the Hamiltonian of the total system, which may be changing due to the changes in the manipulable variables. This process will result in a new thermo-dynamic state  $b = ( \rho_b, H_b )$.   Over the course of the process quantities $\{ Q_i(a \rightarrow b) \}$ of heat may be  exchanged with the reservoirs; the probability distribution over initial conditions, together with the evolution of the joint system, yields a probability distribution over the heats  $\{ Q_i(a \rightarrow b) \}$ .  Let $\ex{Q_i(a \rightarrow b)}_M$ be the expectation value of  the heat obtained from reservoir $B_i$ over the course of the process.  As before, let $\mc{M}(a \rightarrow b)$ be the set of manipulations in $\mc{M}$ that lead from $a$ to $b$.  For any manipulation $M$ in $\mc{M}(a \rightarrow b)$, define
\begin{equation}
\sigma_M(a \rightarrow b) = \sum_i \frac{\ex{Q_i(a \rightarrow b)}_M}{T_i}.
\end{equation}
Define the statistical-mechanical entropy $S_\mc{M}(a \rightarrow b)$ by
\begin{equation}\label{StatEnt}
S_\mc{M}(a \rightarrow b) = \mbox{l.u.b.}\{ \sigma_M(a \rightarrow b) \, | \, M \in \mc{M}(a \rightarrow b) \}.
\end{equation}
We are entitled to use the same notation for this and the entropies as defined in section \ref{EQTD}, as the latter are really only a special case of the entropy defined here, when the probabilities are such that variance in the heat exchanges are negligible.  We are only making explicit the previously implicit dependence on probabilistic considerations.

With these definitions in hand, the statistical-mechanical entropies $S_\mc{M}(a \rightarrow b)$ are defined once we have specified a class of manipulations.  Of particular interest will be classes of manipulations of  the following sort.
\begin{itemize}
\item At time $t_0$, the heat reservoirs $B_i$  have canonical distributions at temperatures $T_i$, uncorrelated with $A$, and are not interacting with $A$.
\item During the time interval $[t_0, t_1]$, the composite system consisting of $A$ and the reservoirs $\{B_i\}$ undergoes Hamiltonian evolution, governed by a time-dependent Hamiltonian $H(t)$,  which may include successive couplings between $A$ and the heat reservoirs $\{B_i\}$.
\item The internal Hamiltonians of the reservoirs $\{B_i\}$ do not change.
\item At time $t_1$, the Hamiltonian of the system $A$ is $H_b$, and, as a result of Hamiltonian evolution of the composite system, the marginal probability distribution of $A$ is $\rho_b$.
\end{itemize}
The initial state of $A$ is arbitrary. No assumption is made about the form of the Hamiltonian $H_A$, the nature of the manipulable variables ${\boldsymbol{\lambda}}$, or about the manipulations applied to them. These could very well include fine-grained manipulations at the molecular level that we would regard as well beyond the range of feasibility.  In what follows, we will use $\mc{M}_\theta$ to designate some class of this sort.  That is, the variable $\mc{M}$ ranges over arbitrary classes of manipulations, and the variable $\mc{M}_\theta$ ranges over classes of manipulations satisfying these conditions.

A class of manipulations of this sort has the advantage that it affords a clear distinction between energy changes of the system $A$ that are to be counted as work, and those that are to be counted as heat.  Changes in energy of $A$ due to manipulation of the exogenous variables  are work; exchanges of energy with the heat reservoirs  are counted as heat.  A more general class of manipulations might include exchanges of energy between the system $A$ and other systems that are not  treated as heat reservoirs---that is, systems with distributions other than canonical distributions. With respect to this class of manipulations, we might not have a neat partition of energy changes to $A$ into heat and work; changes due to interactions with other systems might be classed as neither.

Given some such  class of manipulations,  the second law comes out as a theorem. That is, it  can be proven that
\begin{equation}
S_{\mc{M}_\theta}(a \rightarrow a) \leq 0.
\end{equation}
As we saw in section \ref{EQTD}, it follows from this that if all states are reversibly connectable---that is, if,   for all $a$, $b$,
\begin{equation}\label{rev}
S_{\mc{M}_\theta}(a \rightarrow b \rightarrow a)= 0,
\end{equation}
then there is a state function $S_{\mc{M}_\theta}$, defined up to an arbitrary constant, such that
\begin{equation}
S_{\mc{M}_\theta}(a \rightarrow b) = S_{\mc{M}_\theta}(b) - S_{\mc{M}_\theta}(a).
\end{equation}
If we ask what form that state-function takes, it turns out that, in the classical context, it is  the quantity called the \emph{Gibbs entropy}, and, in the quantum context,  the \emph{von Neumann entropy}.

To show this, we must first define these quantities. Consider  a probability distribution $P$  on a classical state-space $\Gamma$, that has density $\rho$ with respect to Liouville measure. $\rho$ itself may be treated as a random variable: if a point $x$ in $\Gamma$ is randomly selected according to the distribution $P$, there will be a corresponding value of $\rho(x)$. Similarly, any measurable function of  $\rho$ may be treated as a random variable.  We define the Gibbs entropy of the distribution $P$ as proportional to the expectation value, calculate with respect to  $P$, of the logarithm of $\rho$.
\begin{equation}
S_G[\rho] = - k \ex{\log \rho}_P
\end{equation}
For a quantum state, represented by a density operator $\hat{\rho}$, we define the \emph{von Neumann entropy},
\begin{equation}
S_{vN}[\hat{\rho}] =  - k \ex{\log \hat{\rho}}_{\hat{\rho}} = -k \, \mbox{Tr}[\hat{\rho} \log \hat{\rho}].
\end{equation}
Most of what we will have to say applies equally in the classical and quantum contexts.  In what follows, we will use the intentionally ambiguous notation $S[\rho]$ to state results that hold both for  Gibbs entropy of a probability distribution on a classical phase space and for von Neumann entropy of a quantum state.

The link between these quantities and the statistical thermo-dynamic entropy is provided by the following theorem.
\begin{proposition}\label{Fund}\footnote{The classical version of this is found in \citet[pp. 160--164]{GibbsBook}, and the quantum version, in \citet[\S 128--130]{Tolman}.} For any manipulation in the class $\mc{M}_\theta$,
\[
\sum_i \frac{\ex{Q_i}}{T_i} \leq S[\rho_A(t_1)] - S[\rho_A(t_0)].
\]
\end{proposition}Recalling the definition (\ref{StatEnt}) of statistical-mechanical entropies, this gives us,
\begin{proposition}\label{Law2I}  Statistical entropies defined with respect to $\mc{M}_\theta$ satisfy
\[
S_{\mc{M}_\theta}(a \rightarrow b) \leq S[\rho_b] - S[\rho_a].
\]
\end{proposition}
Though not a difficult theorem, Proposition \ref{Fund} is of sufficient importance that it may be called the \emph{Fundamental Theorem of Statistical Thermo-dynamics}. To get a feel for what it  means, consider a heat engine operating in a cycle between a hot heat reservoir at temperature $T_h$ and a cooler heat sink at temperature $T_c$. It extracts a positive amount of heat $Q_h$ from the hot reservoir, performs  work $W$, and discards a positive amount $Q_h - W$ into the sink.  To say that it operates in a cycle means that its initial thermo-dynamic state is restored at the end of this process (it may have built up some correlations with the reservoirs along the way, but these don't matter; the final state is specified by the restriction of the joint probability distribution to the system $A$).  Proposition \ref{Fund} tells us that the expectation values of work obtained,  heat extracted and heat discarded satisfy (recalling that a quantity of heat counts as positive if it is going into the engine and negative if it is going out),
\begin{equation}
\frac{\langle Q_h \rangle}{T_h} - \frac{\ex{Q_h} - \ex{W} }{T_c} \leq 0.
\end{equation}
This gives us, for the expectation value of the work obtained:
\begin{equation}
\langle W \rangle \leq \left(1 - \frac{T_c}{T_h} \right) \langle Q_h  \rangle.
\end{equation}
Thus, the Carnot bound on the efficiency of a cyclical engine operating between these two reservoirs becomes a bound on expectation value of work obtained.  It should be stressed that we have \emph{not} presumed that the actual values of heat exchanges will be or even will probably be close to their expectation values.  No assumption has been made that the probability distributions for these quantities are tightly focussed near the expectation values.  These expectation values satisfy the given relations even if the variance of their distributions is large.

From Proposition \ref{Law2I} the second law of thermo-dynamics is an immediate corollary.
\begin{corollary}\label{Law2II}
For manipulations $\mc{M}_\theta$,
\[
S_{\mc{M}_\theta}(a \rightarrow a) \leq 0
\]
for any thermo-dynamic state $a$.
\end{corollary}
Another immediate corollary of Proposition \ref{Law2I} is,
\begin{corollary}\label{Cor1}
If $S_{\mc{M}_\theta}(a \rightarrow b \rightarrow a)= 0$, then
\[
S_{\mc{M}_\theta}(a \rightarrow b) = S[\rho_b] - S[\rho_a].
\]
\end{corollary}
Thus, the state function whose existence is guaranteed by the second law plus reversibility is, up to an additive constant, the Gibbs or von Neumann entropy.\footnote{It should be stressed that we are not \emph{defining} the statistical mechanical entropy $S_\mc{M}(a \rightarrow b)$ in terms of $S[\rho_b]$ and $S[\rho_a]$; it is defined by (\ref{StatEnt}).}

A probability distribution may encode a lot of details about the microstate of the system that are irrelevant to the results of available manipulations.  Consider, for example, a gas consisting of a macroscopic number of molecules  initially confined to the left side of a container. A partition is removed, and the gas is allowed to expand freely into the whole volume of the container.  Imagine (as is common in the literature on the philosophy of statistical mechanics) that it can do so while isolated from its environment.  Any probability distribution with support in the set of states in which all molecules are one side will evolve into a distribution with support on a set that is a minuscule fraction of the available phase space.  However, this set will so finely distributed that only very fine-grained manipulations could distinguish this probability distribution from an equilibrium distribution uniform in the accessible region of phase space.  If the only available manipulations involve pistons and couplings to heat reservoirs, there will be no difference, in terms of expected reactions to these manipulations, between a probability distribution corresponding to a recent isolated expansion from one side of the box and one on which the gas had been in equilibrium with a heat reservoir for a long time.  The considerable knowledge about the state of the gas that comes from knowing it was in the left half of the box an hour ago is irrelevant to results of ham-handed interventions.

With these considerations in minds, we  define an equivalence-relation between thermo-dynamic states.
\begin{quote}
Any two thermo-dynamic states $(\rho, H_{\boldsymbol{\lambda}})$, $(\rho', H_{\boldsymbol{\lambda}})$  having  the same values of the manipulable variables ${\boldsymbol{\lambda}}$, are \emph{thermo-dynamically equivalent} with respect to $\mc{M}$ if and only if, for every manipulation $M \in \mc{M}$, $\rho$ and $\rho'$  yield the same expectation values for work, $\ex{W}$,   and for heat exchanges, $\ex{Q_i}$,  over the course of the manipulation $M$.  We will write $a \sim_\mc{M} a'$ for thermo-dynamic equivalence.
\end{quote}
We could, of course, define a stronger notion on which equivalence requires, not just equality of expectation values, but equality of the probability distributions for work and heat,  but at the moment I see no need for this.  One could also relax the condition a bit, and require, not exact equality, but equality within a certain tolerance (in which case the relation will not be strictly speaking an equivalence relation).

Define coarse-grained entropies,
\begin{equation}
\bar{S}_\mc{M}[a] = \mbox{l.u.b} \: \{S[a'] \,| \, a \sim_{\mc{M}} a'   \}.
\end{equation}
Obviously, for any state $a$,
\begin{equation}
\bar{S}_\mc{M}[a] \geq S[a].
\end{equation}
If, for some thermo-dynamic state $a$, there is another state $a'$ that is thermo-dynamically equivalent to it and which maximizes the entropy among states equivalent to $a$, we will say that $a'$ is a \emph{coarse-graining} of $a$.   We will say that $a$ is a \emph{coarse-grained state} if and only if $\bar{S}_\mc{M}[a] = S[a]$. Note, however, that the coarse-grained entropy is well-defined whether or not for every state there is a corresponding coarse-grained state.

With the concept of coarse-grained entropy in hand, we have a strengthening of Proposition \ref{Law2I}.
\begin{proposition}\label{Law2III}
For any class of manipulations $\mc{M}_\theta$, and any pair of thermo-dynamic states $a$, $b$,
\[
S_{\mc{M}_\theta}(a \rightarrow b) \leq S[\rho_b] - \bar{S}_{\mc{M}_\theta}[\rho_a].
\]
\end{proposition}

The upper bound on  $S_{\mc{M}_\theta}(a \rightarrow b)$ in Proposition \ref{Law2III} is a difference between two different state-functions, $S$ and $\bar{S}_{\mc{M}_\theta}$, depending on whether the state is the initial or final state of the manipulation. We may call $\bar{S}_{\mc{M}_\theta}$ the \emph{departure entropy}, and $S$,  the \emph{arrival entropy}.

This sheds light on a move that has routinely been made, since the time of Gibbs: the use of a coarse-grained entropy (usually obtained via a coarse-graining of the state) to track approach to equilibrium of an isolated system.  If a system is isolated, the Gibbs/von Neumann entropy is a constant of the motion.  The state can, however, evolve into a state in which the result of any manipulation would be the same as  would obtain if the state were one with a higher entropy $\bar{S}_{\mc{M}_\theta}$.  The quantity $\bar{S}_{\mc{M}_\theta}$, rather than $S$, is the one relevant to bounds on the value of the state for obtaining work, and so is the relevant quantity to track, if one is interested in tracking loss of such value as the system  approaches equilibrium.  This is not, as some have suggested, an \emph{ad hoc} move that is made for the sole purpose of finding a quantity that increases on the way to equilibrium.

From the second law, Corollary \ref{Law2II}, for any $a$, $b$, $S_{\mc{M}_\theta}(a \rightarrow b \rightarrow a)$ cannot be positive.  It follows from Proposition (\ref{Law2III}) that the difference between the Gibbs/von Neumann entropies of the states $a$ and $b$, and the corresponding coarse-grained versions, puts a bound on how close to zero $S_{\mc{M}_\theta}(a \rightarrow b \rightarrow a)$ can be.
\begin{corollary}\label{diss1}
\[
-S_{\mc{M}_\theta}(a \rightarrow b \rightarrow a) \geq \left(\bar{S}_{\mc{M}_\theta}(\rho_a) - S[\rho_a]\right) + \left(\bar{S}_{\mc{M}_\theta}(\rho_b) - S[\rho_b]\right).
\]
\end{corollary}
An immediate consequence of this is that only coarse-grained states can be reversibly connected.
\begin{corollary}
If $S_{\mc{M}_\theta}(a \rightarrow b \rightarrow a)= 0$, then  $\bar{S}_{\mc{M}_\theta}[a] = S[a]$ and  $\bar{S}_{\mc{M}_\theta}[b] = S[b]$.
\end{corollary}

We can summarize the relations between the thermo-dynamic entropies $S_{\mc{M}_\theta}(a \rightarrow b)$ and the Gibbs/von Neumann entropies as follows.
\begin{enumerate}
\item If the states $a$ and $b$ can be connected reversibly, then the thermo-dynamic entropy $S_{\mc{M}_\theta}(a \rightarrow  b)$ is equal to the difference of  the Gibbs/von Neumann entropies of the two states.  That is,
    \[
    S_{\mc{M}_\theta}(a \rightarrow  b) = S[\rho_b] - S[\rho_a].
    \]
    This is not an arbitrary or whimsical choice, but a theorem.
\item This relation between thermo-dynamic entropy and the Gibbs/von Neumann entropy can hold for both $S_{\mc{M}_\theta}(a \rightarrow  b)$ and $S_{\mc{M}_\theta}(b \rightarrow  a)$ \emph{only if} $a$ and $b$ can be connected reversibly.  If they cannot, then either $S_{\mc{M}_\theta}(a \rightarrow  b)$ is strictly less than $S[\rho_b] - S[\rho_a]$, or  $S_{\mc{M}_\theta}(b \rightarrow  a)$ is strictly less than $S[\rho_a] - S[\rho_b]$ (or both).
\item If $a$ is not a coarse-grained state, then  $S_{\mc{M}_\theta}(a \rightarrow b)$ is never equal to $S[b] - S[a]$ for any state $b$ that can be reached from $a$, but is always strictly less.
\end{enumerate}

To get a feel for this, suppose that $a$ and $b$ can be connected adiabatically, that is, purely Hamiltonian evolution can take $\rho_a$ to $\rho_b$.  One can think of free expansion of an adiabatically isolated gas; $\rho_b$ is then  distribution that has support on a small but highly fibrillated set that is stretched out throughout the available phase space.  Then, because Hamiltonian evolution preserves $S$, $S[\rho_b]$ is equal to  $S[\rho_a]$.  It would simply be a gross error to conclude from this that $a$ and $b$ are entropically on a par, and that, for some state $c$ that can be reached from both, $S_\mc{M}(a \rightarrow c)$ is equal to $S_\mc{M}(b \rightarrow c)$.  Unless the expansion can be undone adiabatically (which would require fantastically fine-grained control over the evolution of the system), $S_\mc{M}(b \rightarrow c)$ is strictly less than $S_\mc{M}(a \rightarrow c)$.

\section{Dissipation} In any process $M$ that takes a state $a$ to a state $b$, some of the work done, or heat discarded into a reservoir, may be recovered by some process that takes $b$ back to $a$.  If the process can be reversed with the signs of all $\ex{Q_i}$ reversed, then full recovery is possible.  If full recovery is not possible, and cannot even be approached arbitrarily closely, we will say that the process is \emph{dissipatory}.  A manipulation $M'$ that takes $b$ to $a$ and recovers work done and heat discarded would be one such that
\begin{equation}
\sigma_M(a \rightarrow b) + \sigma_{M'}(b \rightarrow a) = 0.
\end{equation}
There might be a limit to how closely this can be approached. Define the dissipation associated with the process of $M$ taking $a$ to $b$ as the distance between this limit and perfect recovery.
\begin{align}\label{dissdef}
\nonumber \delta_M(a \rightarrow b) &= \mbox{g.l.b.}\{ -(\sigma_M(a \rightarrow b) + \sigma_{M'}(b \rightarrow a)) \: | \, M' \in \mc{M}(b \rightarrow a) \}
\\ &= -S_\mc{M}(b \rightarrow a) - \sigma_M(a \rightarrow b).
\end{align}
It follows from the  second law that this is non-negative.

If there is no limit to how much the dissipation associated with processes that connect $a$ to $b$ can be diminished, $S_\mc{M}(a \rightarrow b \rightarrow a)$ is equal to zero.  This is the condition that we earlier called reversibility.  It is easy to see that the negative of this places a bound on the minimal dissipation associated with any manipulation that takes $a$ to $b$.   For any $M$ in $\mc{M}(A \rightarrow b)$,
\begin{equation}
\delta_{\mc{M}}(a \rightarrow b) \geq -S_{\mc{M}}(a \rightarrow b \rightarrow a).
\end{equation}
It follows from this and Corollary \ref{diss1} that the difference between the coarse-grained and non-coarse grained versions of the Gibbs/von Neumann entropies of the states $a$ and $b$ place bounds on the minimal dissipation associate with a process that takes $a$ to $b$.
\begin{corollary}  For any states $a$, $b$, and any manipulation $M$ in $\mc{M}_\theta$,
\[
\delta_M(a \rightarrow b) \geq \left(\bar{S}_{\mc{M}_\theta}(\rho_a) - S[\rho_a]\right) + \left(\bar{S}_{\mc{M}_\theta}(\rho_b) - S[\rho_b]\right).
\]
\end{corollary}

\section{Demonology}\label{demon}  As noted above, in Proposition \ref{NoDemon},  no adiabatic transformation can decrease the entropy of a state.  This is a consequence of the definition of the  entropies $S_\mc{M}(a \rightarrow b)$.  One could also consider transformations of a system $A$ that involve manipulation of $A$ and an auxiliary system $C$ that can couple to it.  No adiabatic transformation can decrease the entropy of the joint system $AC$.

 These  entropies are, of course, defined relative to a class of manipulations.   This dependence  of the question of whether a given process involves an atio increase of entropy on the class of manipulations considered was illustrated by Maxwell via a thought experiment, in which we imagine a ``very observant and neat-fingered being''\footnote{Letter to P. G. Tait, 11 Dec. 1867, in  \citet[p. 332]{MaxLPII}.} capable of performing manipulations that are ``at present impossible to us'' \citep[p. 308]{MaxTOH}.

Suppose we have a class $\mc{M}$ of manipulations, and supplement it with some manipulation not in the class, to form a new class $\mc{M}^+$.    It could happen that some state-transformation effected adiabatically via  manipulations in $\mc{M}^+$ could lower the entropy of a state, relative to $\mc{M}$.  That is, there might be an adiabatic transformation $a \rightarrow b$, achievable via manipulations in $\mc{M}^+$,  such that, for some state $c$, $S_\mc{M}(b \rightarrow c) >  S_\mc{M}(a \rightarrow c)$.  Someone confused about the dependence of entropy on a set of manipulations might take this to be a violation of the principles of thermo-dynamics, which dictate that, if an adiabatic process can take $a$ to $b$, $S_\mc{M}(b \rightarrow c) \leq S_\mc{M}(a \rightarrow c)$.  There is no such violation, because $S_{\mc{M^+}}(b \rightarrow c) \leq S_{\mc{M^+}}(a \rightarrow c)$.

This can be vividly illustrated by imagining a stock $\mc{M}$ of physically possible manipulations to be supplemented by a magical instantaneous velocity-reversing operation, yielding an enhanced set $\mc{M}^+$.  Consider our stock example of a container of gas, and let $\mc{M}$ be the usual sorts of manipulations, consisting of manipulations of the position of the piston and heat exchanges with various heat reservoirs.  Let $\mc{M}^+$ be this stock of operations, supplemented by a magical velocity-reversal.   Consider a container of gas initially confined to a subvolume, which expands to fill the whole container.  With respect to $\mc{M}$, this expansion counts as an entropy increase. An irreversible expansion is a lost opportunity to obtain work.   But, since, with respect to  $\mc{M}^+$, the expansion is adiabatically reversible, there is no entropy increase, no lost opportunity to obtain work, as one can apply the reversal operation and wait for the gas to return to its original subvolume.  An application of the velocity reversal operation to the expanded gas results in an entropy decrease with respect to $\mc{M}$ but not $\mc{M}^+$.  Since the operation preserves phase-space volume (or, in the quantum context, the absolute value of the inner product of any two state-vectors), the proof of Proposition \ref{Fund} still goes through, and the statistical version of the second law holds even for the set $\mc{M}^+$ of manipulations.  A demon capable of performing a velocity reversal could undo the process of equilibration, but could not operate an engine in a cycle to violate the Carnot bound on efficiency of a heat engine.

This may seem paradoxical to some. \emph{Surely}, it will be said,  a gas that is initially spread out throughout a container and subsequently retreats to a corner \emph{must} be decreasing its entropy.  This cannot be sustained, however, if one attends to the definition of thermodynamic entropy.  If the expansion of a gas can be can be reversed adiabatically, then, by the definition of thermodynamic entropy---not just the definition we have given but by the definitions found in all textbooks of thermodynamics---it is not an entropy-increasing process.  The process of returning to the initial subvolume may be a diminution of Boltzmann entropy, but this only illustrates that the connection between Boltzmann entropy and thermodynamic entropy is somewhat tenuous.

\citet{ExorcistI} distinguish between \emph{straight} and \emph{embellished} violations of the second law of thermodynamics.  A straight violation decreases the entropy of an adiabatically isolated system, without compensatory increase of entropy elsewhere.  An embellished violation exploits such decreases in entropy reliably to provide work.  In a similar vein, David \citet{WallaceControlVid}   distinguishes between two types of demon.  Adapting the distinction to our terminology, a demon of the first kind decreases entropy defined with respect to some class $\mc{M}$ of manipulations, by utilizing a manipulation outside the class.  A demon of the second kind violates the Carnot bound on efficiency of a heat engine over a cycle that restores the state of the demon plus any auxiliary system utilized to its original thermo-dynamic state. By Proposition \ref{Fund}, a demon of the second kind cannot exist without a departure from Hamiltonian dynamics.\footnote{It is essential to the theorem that the dynamics preserve phase-space volume.  That this condition is required to underwrite the second law is illustrated by \citet{ExorcistII}, who, building on the work of \citet{Skordos93} and \citet{ZhangZhang}, exhibit a fictitious system with non-Hamiltonian, energy-conserving,  time-reversal invariant dynamics that completely converts heat drawn from a heat reservoir into work.}   A demon of the first kind only illustrates the dependency of entropy on the class of manipulations considered.

Maxwell's purpose in introducing the demon was to illustrate the dependence of thermodynamic concepts on the class of manipulations considered. He was quite explicit  about what the point of the thought-experiment was: to emphasize  the built-in limitation of conclusions drawn from standard thermodynamics to situations in which bodies consisting of a large number of molecules are dealt with in bulk.  These conclusions, he says,  may be found to be inapplicable to situations involving manipulation of individual molecules \citep[pp. 308--309]{MaxTOH}.  Despite this, the point, a fairly simple one, has been widely misunderstood, resulting in  a vast and largely confused literature on the physical possibility or impossibility of a Maxwell demon.

\section{Temporal asymmetry, and thermalization}  The Fundamental Theorem of Statistical Thermo-dynamics, Proposition \ref{Fund}, follows from elementary properties of the Gibbs and von Neumann entropies and of Hamiltonian evolution.  It is  not temporally symmetric.  We consider a transformation that takes state $a$ into state $b$, and the order matters, because the right hand side of the inequality displayed is not invariant under interchange of $a$ and $b$.  No such asymmetry is present in the underlying dynamics.  Where, then, does the temporal asymmetry come in?

The mathematical result on which the Fundamental Theorem   depends is the following (stated here, and proven in the Appendix). Consider a joint system composed of subsystems $A$ and $B$, which undergoes Hamiltonian evolution between times $t_0$ and $t_1$.  The total Hamiltonian $H_{AB}$ may change during the process; changes may be made to $H_A$, corresponding to work done on the system, and to the Hamiltonian of interaction between the two systems.  We assume that at times $t_0$ and $t_1$ the total Hamiltonian is just the sum of the internal Hamiltonians $H_A$ and $H_B$, and that $H_B(t_1) = H_B(t_0)$.  The expectation value of the energy received by $A$ from $B$ is
\[
\ex{Q} = - (\ex{H_B}_{t_1} - \ex{H_B}_{t_0}).
\]
Suppose that the state $\rho_{AB}$ at time $t_0$ is one on which (i) $B$ has canonical distribution $\tau_\beta$, and (ii) $A$ and $B$ are uncorrelated.  The distribution of $A$ at $t_0$ is arbitrary.
\begin{proposition}\label{Law2Sym}   Under the stated conditions,
\[
\frac{\ex{Q}}{T} \leq S[\rho_A(t_1)] - S[\rho_A(t_0)].
\]
\end{proposition}
Proposition \ref{Law2Sym} holds for any Hamiltonian dynamics satisfying the specified conditions, and so does not depend on any time-asymmetry in the underlying dynamics.  In fact, it holds regardless of whether $t_1$ is to the future or past of $t_0$.  The two times do not enter symmetrically into the statement of the theorem, however.  It is assumed that the systems $A$ and $B$ are uncorrelated at $t_0$, and this is not required to hold at $t_1$.  That is the relevant difference between starting point and ending point of the process considered.

It is sometimes said that the rationale for taking the initial state of system + heat reservoir to be one without correlations between them is that this has the status of a default assumption: statistical or probabilistic independence is to be assumed in the absence of any interaction that would create correlations.  This is too quick.  Among the things that can create correlations between systems are events in the common past of two systems.  When we couple a system to a heat reservoir, we are \emph{not} assuming that there are no events in their common past that could potentially lead to correlations.

What we \emph{are} assuming is that the reservoir has thermalized, has undergone a process of equilibration in the course of which details of  its past history, including previous interactions with the rest of the world, have been effectively effaced.  A detailed microdescription might reveal some of these details, but it is expected that these will be irrelevant at the macroscopic scale.  To treat a system as a heat reservoir is to treat the fine details of past interactions it might have had with its environment as irrelevant to its subsequent behaviour.  The task of explaining how and why this happens is an interesting and important one. The process produces thermal systems that the science of \therm can take to be available as resources for manipulations.  The study of equilibration is not, however, the province of \thrm.\footnote{See \citet{ExplainingThermo} for further discussion of these points.}

The  is a tendency to conflate the second law of thermodynamics with the tendency of systems to relax to a state of thermal equilibrium, and this has encouraged the idea that the study of equilibration does fall  within the scope of thermodynamics.  These are not the same thing, however. The distinction can be made vivid by considering the impact on the laws of thermo-dynamics  of a ``Loschmidt demon'' that could magically perform a velocity-reversal.  Such a demon could reverse equilibration of an isolated system, but its operations nevertheless fall within the scope of Proposition \ref{Fund}, and the second law of \therm holds even if the stock of manipulations is expanded to included velocity-reversal.

\section{Conclusion}  The chief differences between the theory whose outlines have been sketched here, which I am calling thermo-dynamics, or \thrm, and the usual textbook presentations of thermodynamics, are two-fold.  One is that we have not assumed that all states are reversibly connectible.  Without this assumption, we do not have available a state-function $S$ such that $S_\mc{M}(a \rightarrow b) = S_\mc{M}(b) - S_\mc{M}(a)$. This is a relatively minor point;  with a little care, it is fairly easy to see that much of thermodynamics goes through without this, and the advantage is that the theory applies in regimes in which the inevitable dissipation involved in every process is not taken to be negligible.

The more important difference is that, whereas the usual treatments say that thermodynamic states are defined relative to a set of variables deemed \emph{macroscopic}, we have defined them in terms of a set of variables deemed \emph{manipulable}.  I maintain that this is the best way to make sense of the usual treatments, and that one will find, if one reads closely, that the relevant variables are indeed being treated as manipulable. For the most part, for the purposes of textbook exposition, as long as attention is confined to the macroscopic domain, and we are not bent on pushing application of the theory into the mesoscopic, it is perfectly acceptable to leave the class of manipulations under consideration implicit. The danger of this, however, is that it might tend to give the impression that entropy is a property of a system, something that it has in and of itself, rather than being defined relative to a class of manipulations.

Whether or not the reader agrees that \therm is the best way to make sense of textbook presentations of thermodynamics and of application of its concepts to the physical world, it should be noncontroversial that it is a legitimate subject. The usual objections to invoking concepts such as manipulability tend to be of two (related) sorts. One is that it brings in excessive subjectivity.  The other is that concepts of that sort are out of place in the study of equilibration.  I hope that I have satisfactorily addressed the former, in the preliminary discussion of manipulability.  The latter is met by a delimitation of scope. Though \therm presumes the availability of systems that can be treated as heat reservoirs, study of the process of thermalization does not fall within its scope.

\section{Acknowledgments}  I am grateful to a number of people with whom I have discussed these matters over the years.  In particular, I thank  Owen Maroney for drawing my attention to what I have called the Fundamental Theorem,  Norton for discussions of reversible processes,  David Wallace for urging the considerations of \S \ref{demon} on me, and Carlo Rovelli for getting me to think more about the concept of manipulability.  I thank also the discussants at the Summer School on Entropy in Split, Croatia, July 2018, and the Southwestern Ontario Philosophy of Physics Reading Group, for helpful questions and feedback.

\section{Appendix}
\subsection{Proof of the Fundamental Theorem of Statistical Thermo-dynamics}
In this appendix we prove Proposition \ref{Fund}.

As before, $\rho$ is used ambiguously for either a density function with respect to Liouville measure   on classical phase space, or a quantum density operator.   Hamiltonian evolution is, in the classical context, evolution according to Hamilton's equations of motion, and, in the quantum context implemented  by a family of unitary operators $U(t)$.  The letter $S$, without subscript, denotes either the Gibbs entropy or the von Neumann entropy.

In the classical context, the salient fact about Hamiltonian evolution---and, indeed, the only fact that we will use---is that Liouville measure is invariant under evolution of that type. As a consequence, the expectation value, with respect to Liouville measure $\Lambda$,  of any measurable function on phase space is invariant under Hamiltonian evolution; this includes in particular the Gibbs entropy
\begin{equation}
S_G[\rho] = -k \ex{\rho \log \rho}_\Lambda.
\end{equation}

In the quantum context, the salient fact about Hamiltonian evolution is that it conserves the inner product of two vectors in Hilbert space.  As a consequence, the trace of any operator is invariant; this includes in particular the  von Neumann entropy
\begin{equation}
S_{vN}[\hat{\rho}] = -k \mbox{Tr}[\hat{\rho} \log \hat{\rho}].
\end{equation}
As conservation of phase space volume (classical) and absolute magnitude of inner product (quantum) are the only features of Hamiltonian evolution used, we could expand our repertoire of operations to include fictitious operations, such as an instantaneous velocity reversal, that  retain these features, and the theorem would still go through.

The relevant facts about the Gibbs and von Neumann entropies are:
\begin{enumerate}
\item \emph{Subadditivity.}  For a composite system $AB$,
\[
S[\rho_{AB}] \leq S[\rho_A] + S[\rho_B],
\]
with equality if and only if the subsystems are probabilistically independent.
\item For any  $T > 0$, let $\beta = 1/kT$.  The canonical distribution $\tau_\beta$ minimizes
\[
\ex{H}_\rho - T S[\rho].
\]
\end{enumerate}

With these facts in hand, the proof of the theorem is easy. For brevity, we will write $S_{AB}(t_0)$ for $S[\rho_{AB}(t_0)]$, \emph{etc.}. We will consider only interactions with a single heat reservoir, as the extension to successive interactions with multiple heat reservoirs is merely a matter of repeated application of the theorem.

The evolution from $t_0$ to $t_1$ does not change the joint entropy $S_{AB}$.  At $t_0$, since the systems are uncorrelated, $S_A + S_B$ is at a minimum for the value of $S_{AB}$ that obtains at both $t_0$ and $t_1$.  Therefore,
\begin{equation}
S_A(t_0) + S_B(t_0) \leq S_A(t_1) + S_B(t_1),
\end{equation}
or,
\begin{equation}\label{deltaS}
\Delta S_A + \Delta S_B \geq 0.
\end{equation}
Since $B$ has canonical distribution $\tau_\beta$ at time $t_0$,
\begin{equation}
\ex{H_B}_{t_0} - T S_B(t_0) \leq  \ex{H_B}_{t_1} - T S_B(t_1),
\end{equation}
or,
\begin{equation}
\Delta \ex{H_B} - T \Delta S_B \geq 0.
\end{equation}

This gives us,
\begin{equation}\label{Q}
\ex{Q} = - \Delta \ex{H_B} \leq -T \Delta  S_B.
\end{equation}
From (\ref{deltaS}),
\begin{equation}
- \Delta S_B \leq \Delta S_A,
\end{equation}
which, combined with (\ref{Q}), yields,
\begin{equation}
\ex{Q} \leq T \Delta S_A,
\end{equation}
or,
\begin{equation}
\frac{\ex{Q}}{T} \leq \Delta S_A,
\end{equation}
which is the desired result.

\subsection{Some quotations from the history of \therm}  The science that I am calling \therm is not a new idea. This understanding of the basic concepts of thermodynamics has been present from the very beginning of the subject.  In this appendix I provide some relevant quotations, with no pretense to exhaustiveness.

\bigskip

\noindent \textbf{Josiah Willard Gibbs} (\citeyear[pp. 228--229]{GibbsEHSI}; in \citealt[pp. 166--167]{GibbsSciPapersI}).  Part of this has already been quoted above; here is a fuller quotation.

\par \quad \par

When we say that when two different gases mix by diffusion as we have supposed, the energy of the whole remains constant, and the entropy receives a certain increase, we mean that the gases could be separated and brought to the same volume and temperature which they had at first by means of a certain change in external bodies, for example, by the passages of a certain amount of heat from a warmer to a colder body. But when we say that when two gas-masses of the same kind are mixed under similar circumstances there is no change of energy or entropy, we do not mean that the gases which have been mixed can be separated without change to external bodies. On the contrary, the separation of the gases is entirely impossible. We call the energy and entropy of the gas-masses when mixed the same as when they were unmixed, because we do not recognize any difference in the substance of the two masses. So when gases of different kinds are mixed, if we ask what changes in external bodies are necessary to bring the system to its original state, we do not mean a state in which each particle shall occupy more or less exactly the same position as at some previous epoch, but only a state which shall be undistinguishable from the previous one in its sensible properties. It is to states of systems thus incompletely defined that the problems of thermodynamics relate.

But if such considerations explain the mixture of gas-masses of the same kind stands on a different footing from the mixture of gas-masses of different kinds, the fact is not less significant that the increase of entropy due to the mixture of gases of different kinds, in such a case as we have supposed, is independent of the nature of the gases.

Now we may say without violence to the general laws of gases which are embodied in our equations suppose other gases to exist than such as actually do exist, and there does not appear to be any limit to the resemblance which there might be between two such kinds of gas. But the increase of entropy due to the mixing of given volumes of the gases at a given temperature and pressure would be independent of the degree of similarity or dissimilarity between them. We might also imagine the case of two gases which should be absolutely identical in all the properties (sensible and molecular) which come into play while they exist as gases either pure or mixed with each other, but which should differ in respect to the attractions between their atoms and the atoms of some other substances, and therefore in their tendency to combine with other substances. In the mixture of such gases by diffusion an increase of entropy would take place, although the process of mixture, dynamically considered, might be absolutely identical in its minutest details (even with respect to the precise path of each atom) with processes which might take place without any increase of entropy. In such respects, entropy stands strongly contrast with energy.
\bigskip

\noindent \textbf{Rudolf Clausius} (\citeyear[p. 32]{Clausius77}).  Responding to P. G. Tait's (unfair) charge that the fact that the possibility of a demon that could, without expenditure of work, cool a body below the temperature of its surroundings  ``is absolutely fatal to Clausius' reasoning,'' (\citealt[pp. 118-120]{TaitLecturesII}; see also \citealt[p. 37]{TaitSketchII}), Clausius wrote,

\par \quad \par

Dieses kann ich in keiner Weise zugeben. Wenn die W\"arme als eine Molecularbewegung betrachtet wird, so ist dabei zu bedenken, dass die Molec\"ule so kleine K\"orpertheilchen sind, dass es f\"ur uns unm\"oglich ist, sie einzeln wahrzunehmen.  Wir k\"onnen daher nicht auf einzelne Molec\"ule f\"ur sich allein wirken, oder die Wirkungen einzelner Molec\"ule f\"ur sich allein erhalten, sondern haben es bei jeder Wirkung, welche wir auf einen K\"orper aus\"uben oder von ihm erhalten, gleichzeitig mit einer ungeheuer grossen Menge von Molec\"ulen zu thun, welche sich nach allen m\"oglichen Richtungen und mit allein \"uberhaupt bei den Molec\"ulen vorkommenden Geschwindigkeiten bewegen, und sich an der Wirkung in der Weise gleichm\"assig betheiligen, dass nur zuf\"allige Verschiedenheiten vorkommen, die den allgemeinen Gesetzen der Wahrscheinlichkeit unterworfen sind. Dieser Umstand bildet gerade die charakteristische Eigenth\"umlichkeit derjenigen Bewegung, welche wir W\"arme nennen, und auf ihm beruhen die Gesetze, welche das Verhalten der W\"arme von dem anderer Bewegungen unterscheiden.

Wenn nun D\"amonen eingreifen, und diese charakteristische Eigen\-th\"umlichkeit zerst\"oren, indem sie unter den Molec\"ulen einen Unterschied machen, und Molec\"ulen von gewissen Geschwindigkeiten den Durchgang durch eine Scheidewand gestatten, Molec\"ulen von anderen Geschwindigkeiten dagegen den Durchgang verwehren, so darf man das, was unter diesen Umst\"anden geschieht, nicht mehr als eine Wirkung der W\"arme ansehen und erwarten,  dass es mit den f\"ur die Wirkungen der W\"arme geltenden Gesetzen \"ubereinstimmt.

\bigskip

This I can in no way concede. If heat is regarded as a molecular motion, it should be remembered that the molecules are  parts of bodies that are so small that it is impossible for us to perceive them individually. We can therefore not act on single molecules by themselves, or obtain effect from individual molecules by themselves, but rather, in every action that we exert on a body or receive from it, we have simultaneously to do with an immensely large collection of molecules, which move in all possible directions and with all the speeds occurring among the molecules, and participate in the action uniformly, in such a way that there occur only random differences, which are subject to the general laws of probability. This circumstance forms precisely the characteristic property of that motion which we call heat, and on it depends the laws that distinguish the behavior of heat from that of other motions.

If now demons intervene, and disturb this characteristic property by distinguishing between the molecules, and molecules of certain speeds are permitted passage through a partition, molecules of other speeds refused passage, then one may no longer regard what happens under these conditions as an action of heat and expect it to agree with the laws valid for the action of heat.

\bigskip

\noindent \textbf{James Clerk Maxwell}

\par \quad \par

Available energy is energy which we can direct into any desired channel.  Dissipated energy is energy we cannot lay hold of and direct at pleasure, such as the energy of the confused agitation of molecules which we call heat.  Now, confusion, like the correlative term order, is not a property of material things in themselves, but only in relation to the mind which perceives them.  A memorandum-book does not, provided it is neatly written, appear confused to an illiterate person, or to the owner  who understands thoroughly, but to any other person able to read it appears to be inextricably confused.  Similarly the notion of dissipated energy could not occur to a being who could not turn any of the energies of nature to his own account, or to one who could trace the motion of every molecule and seize it at the right moment.  It is only to a being in the intermediate stage, who can lay hold of some forms of energy while others elude his grasp, that energy appears to be passing inevitably from the available to the dissipated state (\citeyear[p. 221]{Diffusion}, in \citealt[p. 646]{NivenII}).

\bigskip
The second law relates to that kind of communication of energy which we call the transfer of heat as distinguished from another kind of communication of energy which we call work. According to the molecular theory the only difference between these two kinds of communication of energy is that the motions and displacements which are concerned in the communication of heat are those of molecules, and are so numerous, so small individually, and so irregular in their distribution, that they quite escape all our methods of observation; whereas when the motions and displacements are those of visible bodies consisting of great numbers of molecules moving altogether, the communication of energy is called work.

Hence we have only to suppose our senses sharpened to such a degree that we could trace the motions of molecules as easily as we now trace those of large bodies, and the distinction between work and heat would vanish, for the communication of heat would be seen to be a communication of energy of the same kind as that which we call work. (\citeyear[p. 279]{TaitII}, in  \citealt[p. 669]{NivenII}).

\bigskip  \noindent \textbf{John von Neumann} (\citeyear{vNQuantumErgodic})

\bigskip

If we take into account that the observer can measure only macroscopically then we find different entropy values (in fact, greater ones, as the observer is now
less skilful and possibly can therefore extract less mechanical work from the system) \ldots. (\citealt[p. 214]{vNQuantumErgodicE}, from \citealt[p. 47]{vNQuantumErgodic}).

\bigskip
\noindent \textbf{Harold Grad} (\citeyear[pp. 326--27]{GradManyFaces}).

\par \quad \par

Whether or not a diffusion occurs when a barrier is removed depends not on a difference in physical properties of the two substances but on a decision that we are or are not interested in such a difference (which is what governs the choice of an entropy function) \ldots A very illuminating example is given by the ``spin-echo'' effect. In this experiment, it is found that it is possible to produce a highly ordered microscopic state  and, at a later time, effectively reverse all velocities. To a person who has access to such equipment, a very high level ``reversible'' entropy will be appropriate; to one who has not, a lower order entropy will properly describe all phenomena.

\bigskip

\noindent \textbf{Nicolaas Godfried van Kampen}  (\citeyear[pp. 306--307]{vanKampenGibbs}).   In regards to the difference in expression of entropies for a uniform sample of gas and a system composed of two different gases, van Kampen wrote,

\par \quad \par

The origin of the difference is that two different processes had to be chosen for extending the definition of entropy. They are mutually exclusive; the first one cannot be used for two different gases and the second one does not apply to  a single gas.  But suppose that $A$ and $B$ are so similar that the experimenter has no physical way of distinguishing between them. Then he does not have the semi-permeable walls needed for the second process, but on the other hand the first will look reversible to him.  \ldots The point is, that \emph{this is perfectly justified} and that he will not be led to any wrong results. If you tell him that `actually' the entropy increased when he opened the channel he will answer that this is a useless statement since he cannot utilize the entropy increase for running a machine. The entropy increase is no more physical to him than the one that could be manufactured by taking a single gas an mentally tagging the molecules by $A$ or $B$.

In fact, this still holds when the experimenter would be able to distinguish between $A$ and $B$, by means of a mass spectrograph for instance, but is not interested in the difference because it is not relevant for his purpose. This is precisely what engineers do when they make tables of the entropy of steam, ignoring the fact that it is actually a mixture of normal and heavy water. Thus, whether such a process is reversible or not depends on how discriminating the observer is. The expression for the entropy (which one constructs by one or the other processes mentioned above) depends on whether he  is able and willing to distinguish between the molecules $A$ and $B$. This is a paradox only for those who attach more physical reality to the entropy than is implied by its definition.
\bigskip

\noindent \textbf{Edward T. Jaynes} (\citeyear[p. 5]{JaynesGP}).

\par \quad \par

In the first place, it is necessary to decide at the outset of a problem which macroscopic variables or degrees of freedom we shall measure and/or control; and within the context of the thermodynamic system thus defined, entropy will be some function $S(X_1,\ldots, X_n)$ of whatever variables we have chosen. We expect this to obey the second law $T dS \geq dQ$ only as long as all experimental manipulations are confined to that chosen set. If someone, unknown to us, were to vary a macrovariable
 $X_{n+1}$ outside that set, he could produce what would appear to us as a violation of the second law, since our entropy function $S(X_1,\ldots, X_n)$  might decrease spontaneously, while his $S(X_1,\ldots, X_n, X_{n+1})$ increases.

\bigskip

\noindent \textbf{John Goold, Marcus Huber, Arnau Riera, L\'idia del Rio, and Paul Skrzypczyk} (\citeyear[pp. 1--2]{TopicalReview}).

\par \quad \par
If physical theories were people, thermodynamics would be the village witch. Over the course
of three centuries, she smiled quietly as other theories rose and withered, surviving major
revolutions in physics, like the advent of general relativity and quantum mechanics. The other
theories find her somewhat odd, somehow \emph{different} in nature from the rest, yet everyone
comes to her for advice, and no-one dares to contradict her. Einstein, for instance, called her
`the only physical theory of universal content, which I am convinced, that within the framework
of applicability of its basic concepts will never be overthrown.'

Her power and resilience lay mostly on her frank intentions: thermodynamics has never
claimed to be a means to understand the mysteries of the natural world, but rather a path
towards efficient exploitation of said world. She tells us how to make the most of some
resources, like a hot gas or a magnetized metal, to achieve specific goals, be them moving a
train or formatting a hard drive. Her universality comes from the fact that she does not try to
understand the microscopic details of particular systems. Instead, she only cares to identify
which operations are easy and hard to implement in those systems, and which resources are
freely available to an experimenter, in order to quantify the cost of state transformations.

\newpage
\bibliographystyle{chicago}

\end{document}